\documentclass[sigconf,screen]{acmart}

\usepackage[utf8]{inputenc}
\usepackage{soul}
\usepackage{listings}
\usepackage{hyperref}
\usepackage{enumerate}
\usepackage{booktabs}
\usepackage{caption}
\usepackage{xcolor}
\usepackage{subcaption}
\usepackage{microtype}

\newcounter{observation}
\newcommand{\observation}[1]{\refstepcounter{observation}
	\begin{center}
		\framebox{
			\begin{minipage}{0.93\columnwidth}
				{\bf Summary:} \textit{#1}
			\end{minipage}
		}
	\end{center}
}

\title[One Thousand and One Stories: A Large-Scale Survey of Software Refactoring]{One Thousand and One Stories: \\ A Large-Scale Survey of Software Refactoring}

\begin{document}

\author{Yaroslav Golubev}
\affiliation{
    \institution{JetBrains Research}
    \city{Saint Petersburg}
    \country{Russia}
}
\email{yaroslav.golubev@jetbrains.com}

\author{Zarina Kurbatova}
\affiliation{
    \institution{JetBrains Research}
    \city{Saint Petersburg}
    \country{Russia}
}
\email{zarina.kurbatova@jetbrains.com}

\author{Eman Abdullah AlOmar}
\affiliation{
    \institution{Rochester Institute of Technology}
    \city{Rochester}
    \country{United States}
}
\email{eman.alomar@mail.rit.edu}

\author{Timofey Bryksin}
\affiliation{
    \institution{JetBrains Research}
    \institution{Higher School of Economics}
    \city{Saint Petersburg}
    \country{Russia}
}
\email{timofey.bryksin@jetbrains.com}

\author{Mohamed Wiem Mkaouer}
\affiliation{
    \institution{Rochester Institute of Technology}
    \city{Rochester}
    \country{United States}
}
\email{mwmvse@rit.edu}

\begin{abstract}
    Despite the availability of refactoring as a feature in popular IDEs, recent studies revealed that developers are reluctant to use them, and still prefer the manual refactoring of their code. At JetBrains, our goal is to fully support refactoring features in IntelliJ-based IDEs and improve their adoption in practice. Therefore, we start by raising the following main questions. How exactly do people refactor code? What refactorings are the most popular? Why do some developers tend not to use convenient refactoring tools provided by modern IDEs?
    
    In this paper, we investigate the raised questions through the design and implementation of a survey targeting 1,183 users of IntelliJ-based IDEs. Our quantitative and qualitative analysis of the survey results shows that almost two-thirds of developers spend more than one hour in a single session refactoring their code; that refactoring types vary greatly in popularity; and that a lot of developers would like to know more about IDE refactoring features but lack the means to do so. These results serve us internally to support the next generation of refactoring features, as well as can help our research community to establish new directions in the refactoring usability research.
\end{abstract}



\maketitle

\section{Introduction}

\textit{Refactoring}~\cite{Fowler:1999:RID:311424} is traditionally defined as the process of improving the internal code structure without altering its external behavior. Since this practice had been introduced to a wide audience of software engineers, it has become a crucial tool to maintain high-quality software and to reduce its technical debt. Several refactoring types, involving renaming, moving, and extracting elements have been implemented as actionable tools in modern Integrated Development Environments (IDEs), providing developers with an automatic and safe way to apply these predefined code transformations~\cite{schafer2010specifying,soares2012automated,raychev2013refactoring}. 

Even though all modern IDEs usually have a top-level menu with various options devoted to refactoring, several recent surveys report that
developers are often reluctant to adopt these features and still manually refactor their code~\cite{kim2014empirical,silva2016we}. Despite the high-level maturity of IDEs and the safety and interactivity of their tools, manual refactoring is still widely adopted regardless of its drawbacks and error-proneness. 

While several studies have recently highlighted the lack of automated refactoring usage~\cite{kim2014empirical,hauptmann2015generating,alcocer2020improving,silva2016we,vakilian2014alternate,pinto2013programmers}, 
little is known regarding the reasons hindering the widespread adoption of automated refactoring, especially when it is offered as a built-in feature in modern IDEs. 
Furthermore, existing studies were limited by investigating only a few types of refactoring, as well as by the number of developers surveyed.

The goal of this paper is to share the results of a large-scale survey about refactoring conducted by JetBrains Research, to reflect on what refactorings developers actually use, as well as to raise the awareness of the current usability challenges that developers face when they use the IDE to refactor their code. In particular, we designed a survey of 20 questions to investigate several dimensions related to (1) general background information about respondents, (2) developers' familiarity with refactoring in general, (3) how developers tend to refactor their code, (4) the extent to which they are familiar with the IDE built-in refactoring functionalities, (5) along with their degree of adoption and thoughts on specific refactoring features of IDEs, namely the \textsc{Undo} and \textsc{Preview}. Our survey was sent to paid subscribers of the IntelliJ Platform-based IDEs (IntelliJ IDEA, PyCharm, WebStorm, and others), and we received 1,183 complete responses, achieving a response rate of 6.0\% and a completion rate of 88.9\%.

We provide the refactoring community with a variety of insights that are currently being investigated by JetBrains IDE development teams to plan future enhancements of the refactoring tools. The raw data of the survey is available online.\footnote{Raw data of the survey: \url{https://zenodo.org/record/4923175}} Notably, the survey has revealed the following:

\begin{itemize}

\item While refactoring is a regular and recurrent practice supporting the software
development cycle, two-thirds of respondents confirmed that
their refactoring sessions can take up to an hour or even longer.

\item Some refactorings, such as \textit{Rename} and \textit{Extract} entities, are more popular and intuitive than others, such as \textit{Pull Up} and \textit{Push Down} entities. Despite the existence of built-in refactoring tools, some developers opt for refactoring their code using manual practices: for instance, \textit{Find
and Replace} is popular for \textit{Renaming}, \textit{Copy and Paste} is typically used for \textit{Moving}.

\item The familiarity of developers with IDE refactoring features varies from one refactoring type to another. Also, the adoption of an IDE feature correlates with the popularity of the refactoring itself. For instance, \textit{Rename} refactoring has been found to be the most popular one, and its built-in \textsc{Rename} tool is also heavily used. This is in contrast with \textit{Pull Up / Push Down} refactorings, which are the least popular, and consequently their corresponding tools are used significantly less frequently. 

\item The possibility of \textsc{Preview} and \textsc{Undo} helps in consolidating the usage of refactoring tools, as developers tend to trust code transformations more and understand their impact better. Nevertheless, these functionalities can be considered too complex when developers are performing what they believe to be simple and intuitive refactorings.

\end{itemize}

\section{Related Work}

Some refactoring techniques and formalisms to guarantee program preservation have been reported in an extensive survey study by Mens and Tourwe~\cite{mens2004survey}. The authors discussed the existing literature in terms of refactoring activities and their automation techniques. They reported different types of software artifacts being refactored, along with existing refactoring tool support, and the impact of refactoring in the software process. 
Murphy-Hill \& Black~\cite{murphy2008refactoring} surveyed 112 Agile Open Northwest conference attendees and found that refactoring tools are underused by professional programmers. 
At Microsoft, Kim et al.~\cite{kim2014empirical} surveyed 328 professional software engineers to investigate when and how they do refactoring. When surveyed, the developers mentioned the main benefits of refactoring to be: improved readability (43\%), improved maintainability (30\%), improved extensibility (27\%), and fewer bugs (27\%). When asked what provokes them to refactor, the main reason provided was poor readability (22\%). Only one code smell (\textit{i.e.}, code duplication) was mentioned (13\%). 
Sharma et al.~\cite{sharma2015challenges} surveyed 39 software architects to ask about the problems they are facing whenever they refactor their systems and the limitations of existing refactoring tools they use. Their main findings are: (1) fear of breaking the code restricts developers from adopting refactoring techniques, (2) lack of awareness of the impact of refactoring on code quality is a major obstacle to refactoring tasks, and (3) developers feel reluctant to adopt refactoring because it might result in wasting their resources. 
Oliveira et al.~\cite{oliveira2019revisiting} surveyed 107 developers about their refactoring output, using 7 refactoring types applied to pilot software systems. They found significant differences in the outputs for the same tasks due to the differences in the IDE refactorings they used. In their extended work~\cite{oliveira2020revisiting}, Oliveira et al. confirmed that refactoring implementations of various IDEs (Eclipse, NetBeans, etc.) have differences in all refactoring types. They also reported that these IDEs have different input parameters to apply refactorings.

Our study complements the ongoing effort of previous studies by providing more in-depth insights regarding the challenges faced by developers, specifically when using modern IDEs. 
Our study is also the first to attract 1,183 developers, becoming the largest refactoring survey in literature. 

\begin{table*}[t]
\footnotesize
\centering
\caption{The summary of survey questions. Grey numbers near certain questions indicate that the presence of this question conditionally depends on another question, specified by the number}
\label{fig:survey}
\begin{tabular}{@{}lll@{}}
\toprule
\multicolumn{2}{c}{\textbf{I. Background}}                                                                                                                                                                                                                                                            &  \\ 
\textbf{Question 1}  & How many years of coding experience do you have?                                                                                                                                                                                                                               &  \\
\textbf{Question 2}  & What programming languages do you regularly use?                                                                                                                                                                                                                               &  \\
\midrule
\multicolumn{2}{c}{\textbf{II. Familiarity with refactorings}}                                                                                                                                                                                                                                        &  \\ 
\textbf{Question 3}  & In the past month, how often have you performed any code refactoring?                                                                                                                                                                                                                    &  \\
\textbf{Question 4 \textcolor{gray}{(3)}}  & Have you renamed anything in your code or project structure in the past month?                                                                                                                                                                                                          &  \\
\textbf{Question 5 \textcolor{gray}{(3)}}  & Have you moved any code from one location in the project to another?                                                                                                                                                                                                  &  \\
\textbf{Question 6 \textcolor{gray}{(3)}}  & During this time, did you ever refactor code for an hour or more in a single session?                                                                                                                                                                                                     &  \\
\midrule
\multicolumn{2}{c}{\textbf{III. Refactoring approaches}}                                                                                                                                                                                                                                              &  \\
\textbf{Question 7}  & \begin{tabular}[c]{@{}l@{}}For the following scenarios, please select all the approaches you have used in the past month.\\ (\textit{Renaming a class, method, variable, or symbol / Extracting a method or a variable from existing code / Moving code to another file})\end{tabular}                    &  \\
\textbf{Question 8 \textcolor{gray}{(7)}}  & In these scenarios, what were your main reasons for not using the IDE refactoring feature?                                                                                                                                                                                     &  \\
\textbf{Question 9}  & \begin{tabular}[c]{@{}l@{}}For the following scenarios, please select all the approaches you have used in the past month.\\ (\textit{Inlining a variable or method / Changing the signature of an existing function / Moving a method up or down the class hierarchy})\end{tabular}              &  \\
\textbf{Question 10 \textcolor{gray}{(9)}} & In these scenarios, what were your main reasons for not using the IDE refactoring feature?                                                                                                                                                                                     &  \\
\midrule
\multicolumn{2}{c}{\textbf{IV. IDE refactoring features}}                                                                                                                                                                                                                                             &  \\
\textbf{Question 11} & \begin{tabular}[c]{@{}l@{}}How familiar are you with the following IDE refactoring features?\\ (\textsc{Rename file, class, method, symbol, etc.} / \textsc{Extract method, variable, component, etc.} / \textsc{Move} / \\\textsc{Inline variable or method} / \textsc{Change signature} / \textsc{Pull Up or Push Down member})\end{tabular} &  \\
\textbf{Question 12 \textcolor{gray}{(11)}} & Please think about the last several times you used IDE refactorings. How happy were you with the overall experience?                                                                                                                                                   &  \\
\textbf{Question 13 \textcolor{gray}{(12)}} & Please tell us a bit more about your experience.                                                                                                                                                                                                                               &  \\
\textbf{Question 14 \textcolor{gray}{(11)}} & How often do you undo or revert an IDE refactoring action because you’re unhappy with the result?                                                                                                                                                                              &  \\
\textbf{Question 15 \textcolor{gray}{(14)}} & The last few times you undid or reverted an IDE refactoring feature, what were the reasons?                                                                                                                                                                                    &  \\
\midrule
\multicolumn{2}{c}{\textbf{V. Previewing refactorings}}                                                                                                                                                                                                                                               &  \\
\textbf{Question 16} & When refactoring the code, do you find it useful to preview all of your changes before applying them?                                                                                                                                                                          &  \\
\textbf{Question 17 \textcolor{gray}{(16)}} & For what types of changes do you find this feature most useful?                                                                                                                                                                                                                &  \\
\textbf{Question 18 \textcolor{gray}{(16)}} & What are your main reasons for wanting to see a refactoring preview?                                                                                                                                                                                                           &  \\
\textbf{Question 19 \textcolor{gray}{(16)}} & What are your main reasons for not finding a preview useful?                                                                                                                                                                                                                   &  \\
\midrule
\multicolumn{2}{c}{\textbf{VI. Final thoughts}}                                                                                                                                                                                                                                                       &  \\
\textbf{Question 20} & Please share any thoughts or feedback you have about using the IDE refactorings.                                                                                                                                                                                               &  \\ \bottomrule
\end{tabular}
\end{table*}

\section{Study design}
\label{sec:study_design}

\subsection{Pilot Survey}

Since this survey was going to be distributed to a wide range of participants, it was critical to ensure that our questions properly convey the points we are seeking answers to. Therefore, we performed a pilot study for the purpose of refining our questions and survey protocol. The pilot version of the survey contained the following questions: (1) What is your experience in software engineering and what programming languages do you regularly use? (2) Provided a list of popular refactorings, please select whether you know each one of them in general and as an IDE feature. (3) For the same refactorings, please select which ones you use and how often. (4) What are your general thoughts about the current state of automatic refactorings? (5) What are your negative experiences with automatic refactorings? (6) If you had cases when you wanted to perform a refactoring, but decided not to, what were the reasons? (7) How often do you use the \textsc{Preview} IDE feature when performing a refactoring? What are your main reasons for using it? (8) How often do you use the \textsc{Undo} action after applying a refactoring? What are your main reasons for using it?

The pilot version of the survey was reviewed by the members of the Product Management team and the members of the Market Research and Analytics team at JetBrains, who are experienced with survey design and execution. We have received the following feedback to refine the questions.

\begin{itemize}
    \item In questions (2) and (3), we immediately show a list of refactoring names to respondents, thus possibly alienating developers who may perform refactorings but do not know them by their names. Instead, we should tune our questions according to the participants' familiarity with refactorings.
    \item It is better to strictly divide questions about refactoring activities in general and questions about IDE refactoring features in order not to confuse less experienced participants.
    \item To avoid generic responses, it is recommended to tie respondents to a specific time frame: asking about their experience during the past month or during one programming session.
    \item Question (6) is too broad, it is better to split it into more specific questions. Question (7) is restricted to participants who \textit{use} previews, whereas it would be also interesting to learn the reasons for not using that feature.
\end{itemize}

This feedback allowed us to reformulate some questions and consolidate others. The resulting survey was then approved by the Product Management team and the Market Research and Analytics team, and its questions are enumerated in Table~\ref{fig:survey}.

\vspace{-0.2cm}

\subsection{Final Survey}
\label{sec:final_survey}

Now, let us describe all the research dimensions encapsulated in the final version of the survey. 

The complete list of questions is presented in Table~\ref{fig:survey}. There was a total of 20 questions in the survey: 9 single choice, 8 multiple choice (in 4 of which a write-in \textit{Other} answer was available), and 3 open-ended. Also, since our survey targeted different groups of developers, it contained \textit{conditional questions} and one case of \textit{branching}. Conditional questions allow us to gather deeper insights and target specific groups of respondents, and branching allows tailoring questions to a specific group in case groups do not intersect (for example, when some developers like a feature, and the others do not). According to the survey methodology and the guidelines proposed by Kitchenham and Pfleeger~\cite{kitchenham2008personal,easterbrook2008selecting,groves2011survey}, conditional questions and branching ensure that respondents are asked only those questions that apply to them while allowing us to gather deeper insights. You can find more information about conditional questions and branching in our supplementary data package~\cite{extended}. 
Let us now briefly overview the main sections of the survey.

\vspace{-0.2cm}

\begin{enumerate}[I]
    \item \textbf{Background}. The first introductory section had just two basic questions related to the respondent's coding experience and the programming languages they use.
    \item \textbf{Familiarity with refactorings}. The second section aimed to determine the respondent's familiarity with refactorings. We asked them how often they have been recently refactoring code. In case the respondents said that they do it rarely or never at all, we followed up by asking how often they have \textit{renamed} or \textit{moved} code elements to ensure that their previous selection was not due to their unfamiliarity with the term \textit{refactoring}. 
    Finally, we asked the respondents whether they have spent one hour or more in a single session refactoring their code.
    \item \textbf{Refactoring approaches}. The third section was dedicated to discovering whether developers actually carry out refactorings with IDE refactoring tools or manually. In the cases when they did not refactor using IDE tools, we asked a follow-up question to inquire about the reasons for that.
    \item \textbf{IDE refactoring features}. In the fourth section, we concentrated on the automatic refactoring tools of IDEs. 
    We asked the respondents how often they use each type of refactoring features (\textit{e.g.}, \textsc{Rename}). If at least one refactoring feature was regularly used, we presented more questions to gauge their satisfaction with these features. We also inquired about how often respondents had to undo an applied refactoring.
    \item \textbf{Previewing refactorings}. The fifth section discussed the usage of previews when refactoring code. We divided respondents into those who find previews useful and those who do not, and then dived deeper into the reasons for both of these views.
    \item \textbf{Final thoughts}. Finally, in the last section of the survey, we gave the respondents the opportunity to share any refactoring-related feedback.
\end{enumerate}

\subsection{Surveying Process}

We selected respondents from the list of paid subscribers of IntelliJ-based IDEs (IntelliJ IDEA, PyCharm, CLion, WebStorm, etc.)~\cite{intellij} who had previously agreed to receive invitations to such surveys by email. In total, the mailing list consisted of 20,000 developers, compiled evenly from users of different IntelliJ-based IDEs. This allowed us to mitigate possible skews towards a given IDE or specific languages and inquire about broader aspects of refactoring.

19,860 emails were successfully delivered, respondents were given three weeks to fill the survey. In the end, we received 1,330 responses, out of which 1,183 were complete. This corresponds to a \textbf{response rate} of 6.0\% and a \textbf{completion rate} of 88.9\%. We considered the obtained response rate to be satisfactory, as it is close to other studies in the field that had reported a response rate between 5.7\%~\cite{PassosTSE} and 7.9\%~\cite{FlavioMedeiros_2018}

The first two questions in the survey gave us some background information on the developers. The question about coding experience (\textbf{Question 1}) demonstrated that 27.8\% of the participants have more than 16 years of experience, a total of 48.1\% have more than 10 years of experience, and as much as 79\% have more than 5 years of experience. Only 2 respondents (0.2\%) in our survey answered \textit{I~do not have any coding experience}, and were therefore taken to the end of the survey.

\textbf{Question 2} allowed our participants to select up to 3 languages that they are familiar with. The most popular languages by the percentage of respondents in our survey are: JavaScript (46.6\%), Python (28.3\%), PHP (27.2\%), Java (24.4\%), and SQL (21.4\%).
These languages constitute some of the most popular programming languages used in different software engineering domains~\cite{developer_report}. 
Among other languages selected by the participants are Typescript, C/C++, C\#, Go, Ruby, and others. 

Overall, the results of this survey represent the opinions of more than a thousand developers with a high average experience in different popular programming languages, which can serve as a practical window to the current general state of refactoring usage.

\section{Results}

In this section, we describe the results of the survey, divided into five parts, corresponding to sections \textbf{II}--\textbf{VI} mentioned in Section~\ref{sec:final_survey} and shown in Table~\ref{fig:survey}. 

Because of the conditional questions and branching, questions in our survey have a different number of respondents answering them. To avoid creating confusion with numbers not summing up, we decided to present the results of all questions in the form of percentages with respect to the number of respondents who answered them. For all survey questions, the number of respondents is explicitly stated in captions of the figures and tables. Also, to avoid confusion in similar terms, we mark \textit{refactorings in general} in cursive and mark \textsc{IDE refactoring features} in small caps.

To process the open-ended questions, we used the open coding technique based on guidelines provided by Cruzes et al.~\cite{cruzes2011recommended}. For each question, firstly, the list of possible categories of answers was compiled. This was done independently by the first two authors in two iterations. On the first iteration, each author studied the responses and gave them possible labels, and on the second iteration, each author reduced the overlap between the labels and drew up the final list of categories. After this, the first two authors compared their lists of obtained categories and compiled the final list together. Finally, using this final list, they independently did a third iteration and labeled each response (no category, one specific category, or several categories in the case of long responses). After this, they compared and discussed the resulting labeling. In all the questionable cases, the authors had a discussion, until a perfect consensus was reached. This was done for all open-ended questions.

\subsection{Familiarity with Refactorings}

\begin{figure}[h]
  \centering
  \includegraphics[width=2.7in]{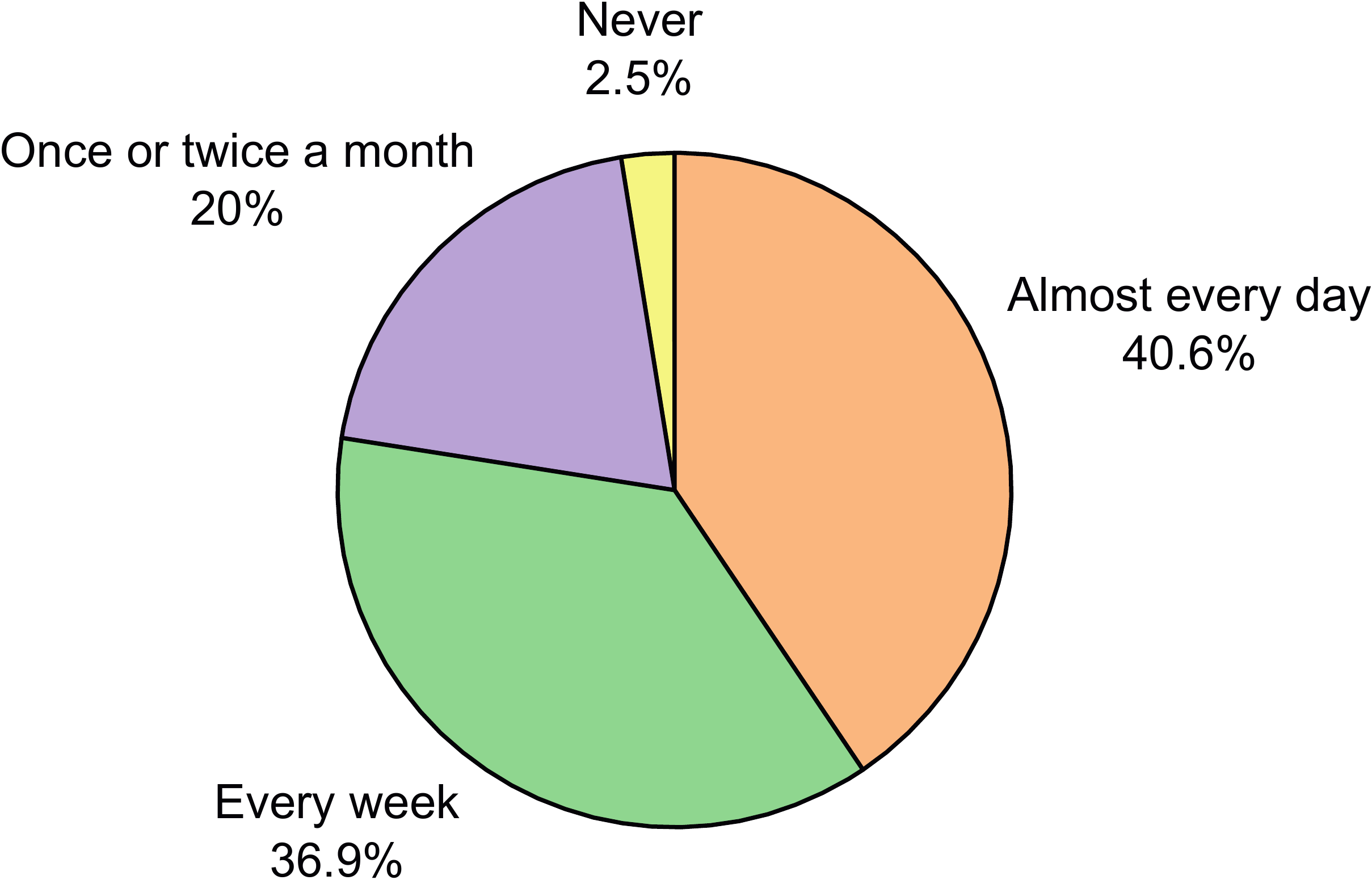}
  \caption{The answers to \textbf{Question 3}: \textit{In the past month, how often have you performed any code refactoring?} (Out of 1,181 respondents)}
  \label{fig:question3}
  \vspace{-0.3cm}
\end{figure}

We started the main body of the survey with the basic \textbf{Question 3} that asked how often the respondents performed any code refactoring recently. Figure~\ref{fig:question3} shows the breakdown of the answers. It can be seen that refactoring is an omnipresent practice in software development: 40.6\% of developers indicated that they refactor code \textit{Almost every day} and 36.9\% more said that they refactor code \textit{Every week}. Only about 20\% of users said that they refactor code \textit{Once or twice a month}, and just 2.5\% \textit{Never} refactored code.

However, it might be the case that developers may have a different perception of the word \textit{refactor}. To account for this, if the participant selected \textit{Once or twice a month} or \textit{Never} in the previous question (\textbf{266} respondents), we additionally showed them \textbf{Questions 4} and \textbf{5}, asking whether they renamed or moved anything in the code recently. We found out that 84.2\% of these 266 respondents renamed entities in the code and 78.2\% of them moved code elements. Overall, only 14 respondents did not answer positively to either of these questions, and therefore, we did not consider their answers further, because they are not a target audience for any questions about refactorings. The remaining survey considered the answers of all the remaining 1,167 participants.

\begin{figure}[h]
  \centering
  \includegraphics[width=1.9in]{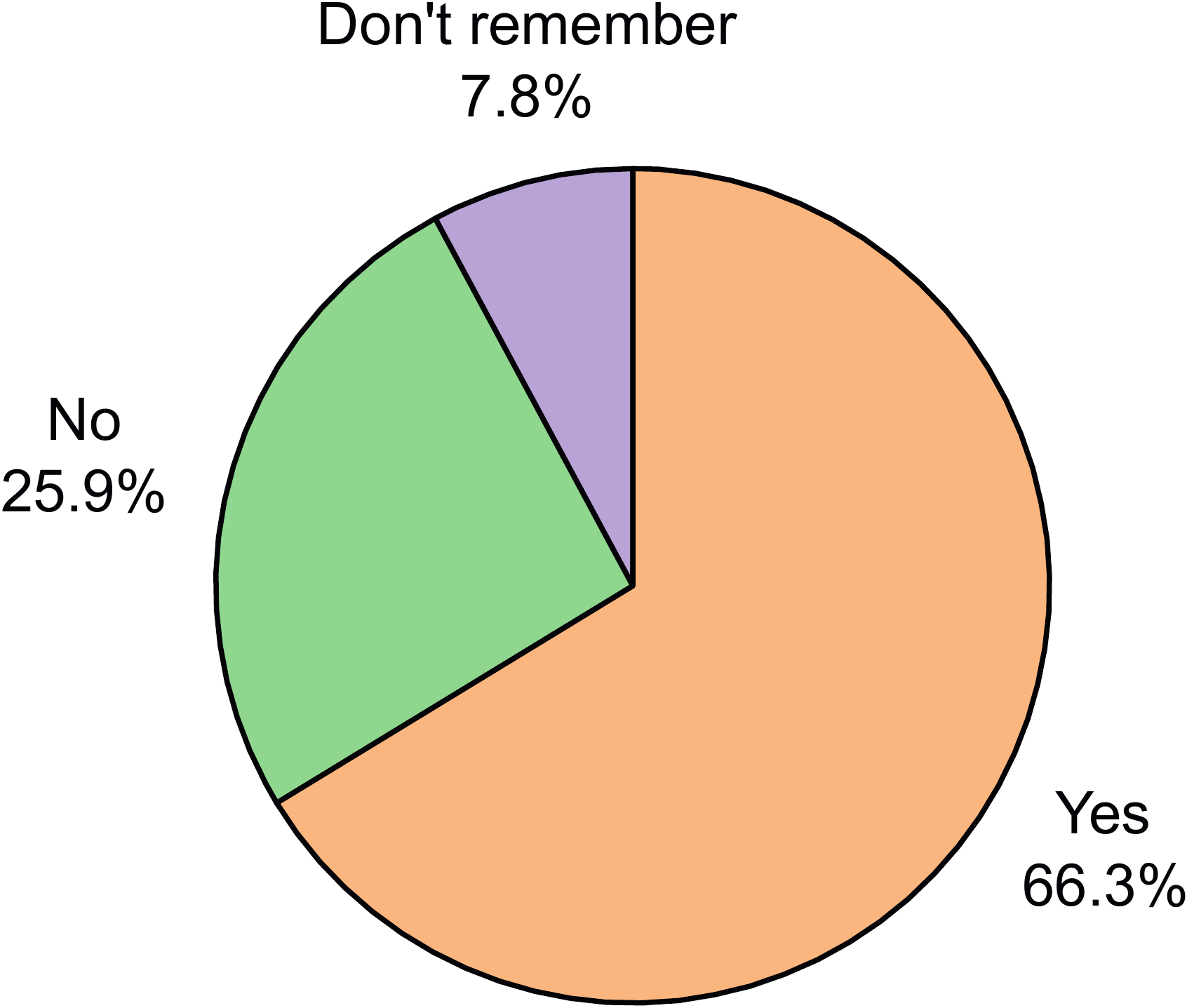}
  \caption{The answers to \textbf{Question 6}: \textit{During this time, did you ever refactor code for an hour or more in a single session?} (Out of \textbf{1,145} respondents)}
  \label{fig:question6}
  \vspace{-0.2cm}
\end{figure}

Respondents who are familiar with refactoring (\textbf{1,145} respondents) were exposed to the last question in this block, \textbf{Question 6}, asking whether they have been recently refactoring for an hour or more in a single session. The results are presented in Figure~\ref{fig:question6}. Surprisingly, almost \textit{two-thirds} of developers answered positively. 

While it is widely accepted that refactoring helps to enforce better design practices or to cope with design defects~\cite{Fowler:1999:RID:311424}, recent studies have shown that developers interleave refactoring activities with other maintenance and development-related tasks in practice, including feature updates, bug fixes, and API types migration~\citep{tsantalis2018accurate,ketkar2020understanding,alomar2021refactoring,pantiuchina2020developers,alomar2021we,murphy2011we}. Yet, little is known about the overhead that refactorings exhibit, especially when they are executed in conjunction with these development tasks. According to the survey results, the majority of developers state that they spent over an hour while refactoring their code, which is interesting, considering that refactoring tools were designed to be executed independently, applying small edits.
This finding was interesting to us, and we plan to monitor how refactoring features are being used for such time frame in the future. Researchers should also pay close attention to such refactoring sessions and study whether our current refactoring tools remain efficient during continuous refactoring.

\vspace{0.1cm}
\observation{Intuitively, refactorings are a key element in the software development cycle, according to the participants of the survey. Nearly four out of five developers indicated that they refactored code every week or even almost every day recently. Interestingly, two-thirds of respondents said that they had refactoring sessions of an hour or longer during this time.}

\subsection{Refactoring Approaches}
\label{sec:approaches}

\begin{figure*}
\begin{subfigure}{3.35in}
  \centering
  \includegraphics[height=2.1in]{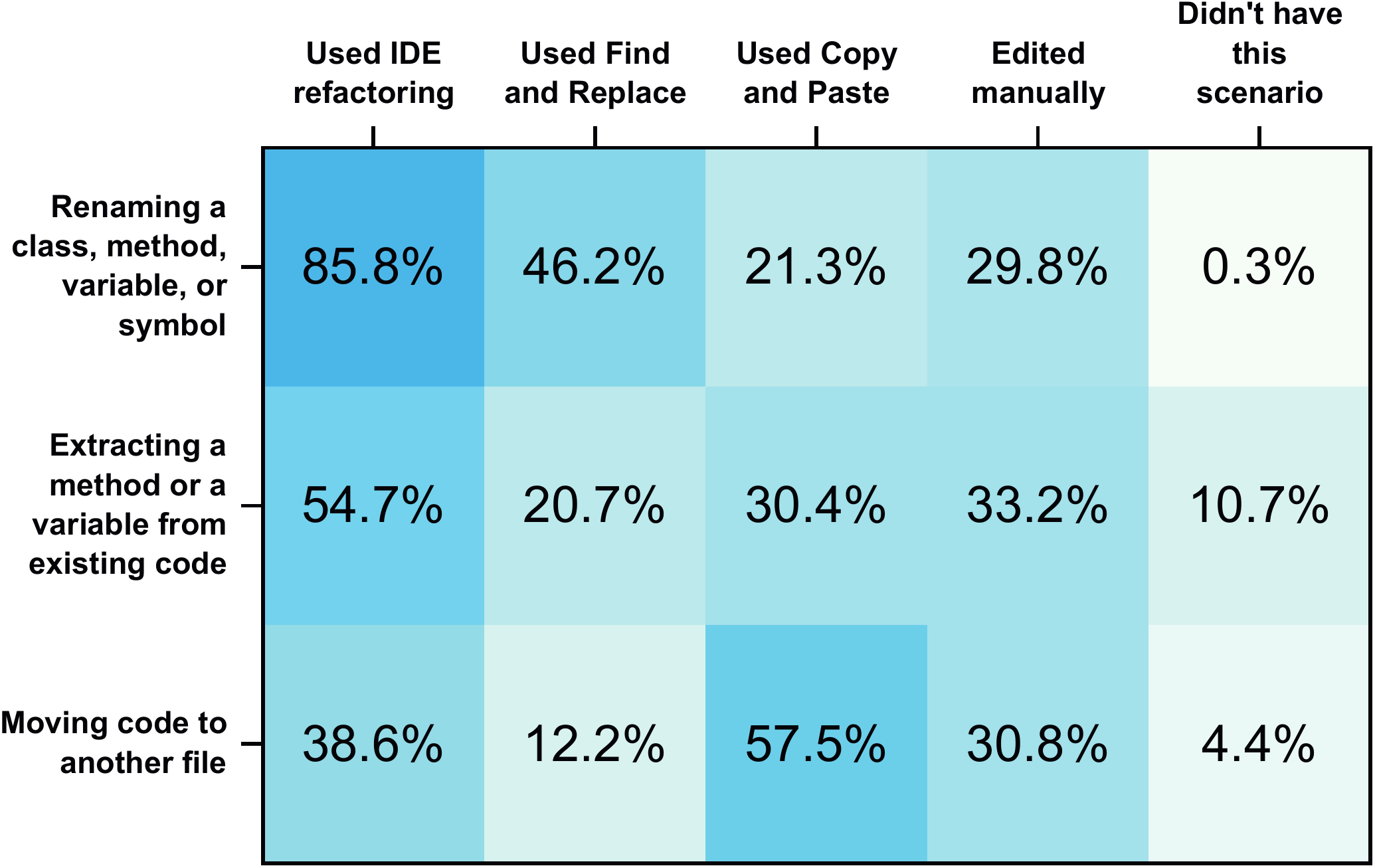}
  \caption{The answers to \textbf{Question 7}: \textit{For the following scenarios, please select all the approaches you have used in the past month.} Several approaches can be selected for each refactoring, the percentages show the proportion of respondents who chose each option. (Out of 1,167 respondents)}
  \label{fig:question7}
\end{subfigure}
\hspace{0.2cm}
\begin{subfigure}{3.35in}
  \centering
  \includegraphics[height=2.3in]{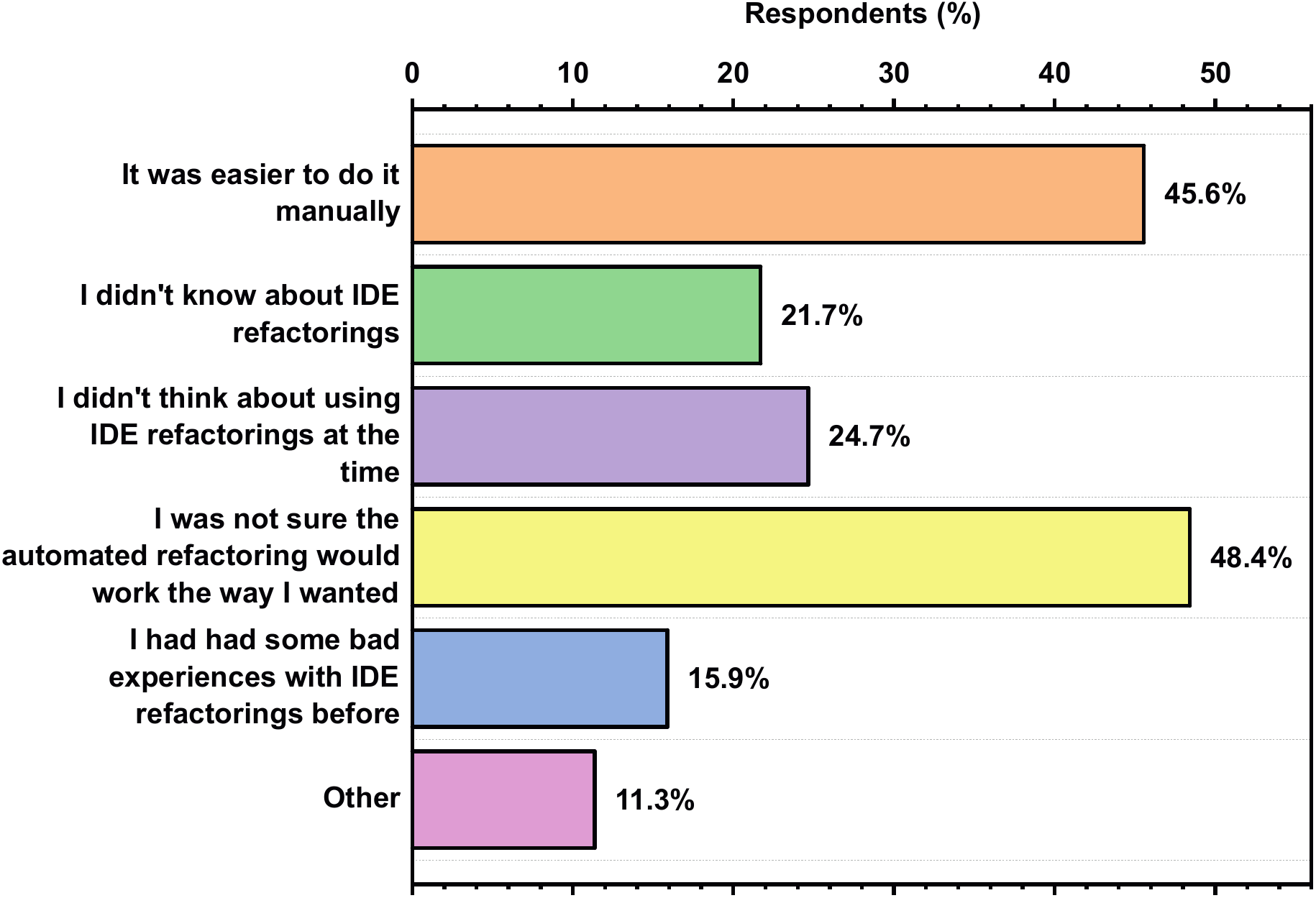}
  \caption{The answers to \textbf{Question 8}: \textit{In these scenarios (Renaming, Extracting, Moving), what were your main reasons for not using the IDE refactoring feature?} (Out of 1,014 respondents)}
  \label{fig:question8}
\end{subfigure}
\newline
\begin{subfigure}{3.35in}
  \centering
  \hspace*{-0.2cm}\includegraphics[height=2.1in]{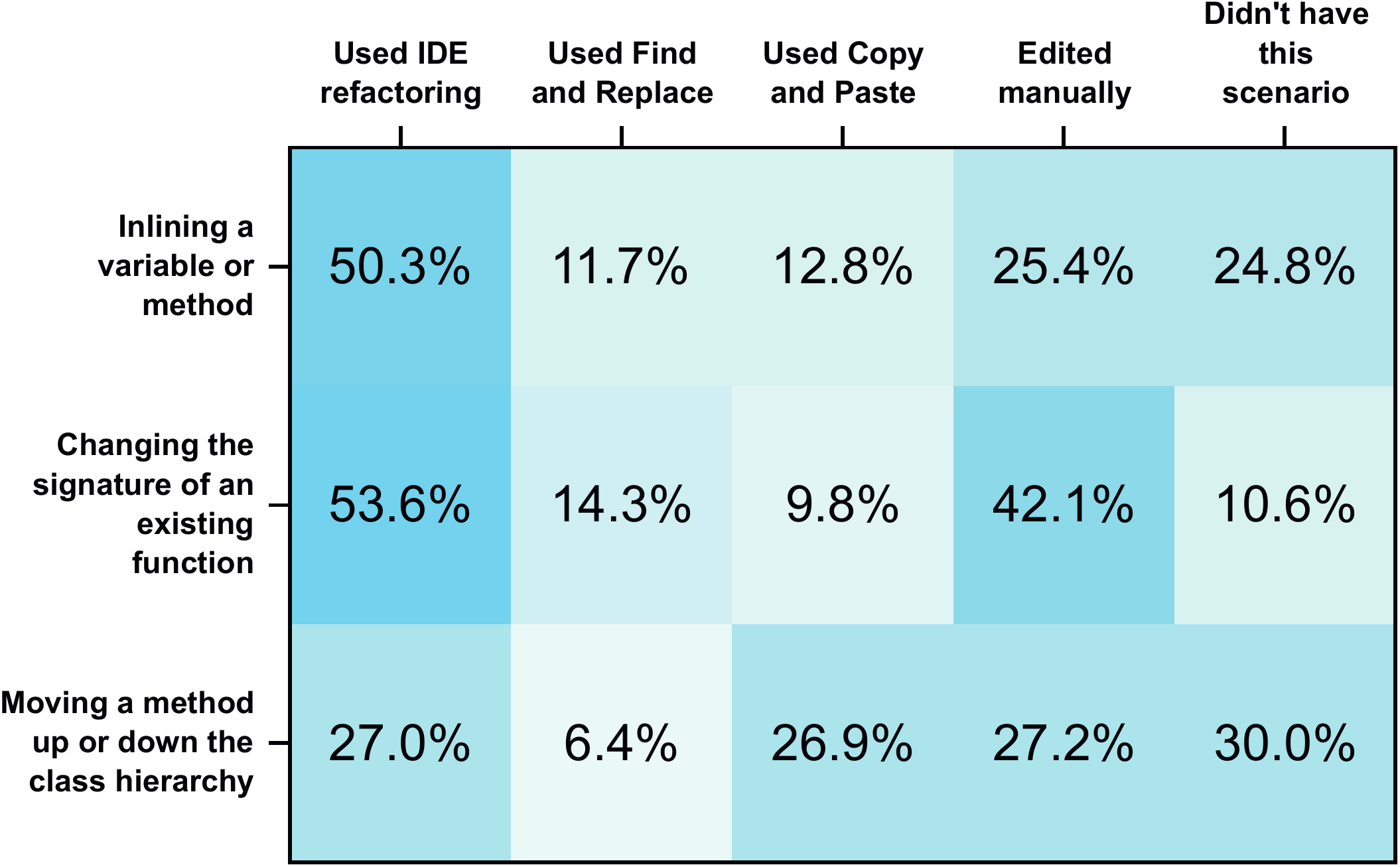}
  \caption{The answers to \textbf{Question 9}: \textit{For the following scenarios, please select the approaches you’ve used in the past month.} Several approaches can be selected for each refactoring, the percentages show the proportion of respondents who chose each option. (Out of 1,167 respondents)}
  \label{fig:question9}
\end{subfigure}
\hspace{0.2cm}
\begin{subfigure}{3.35in}
  \centering
  \includegraphics[height=2.3in]{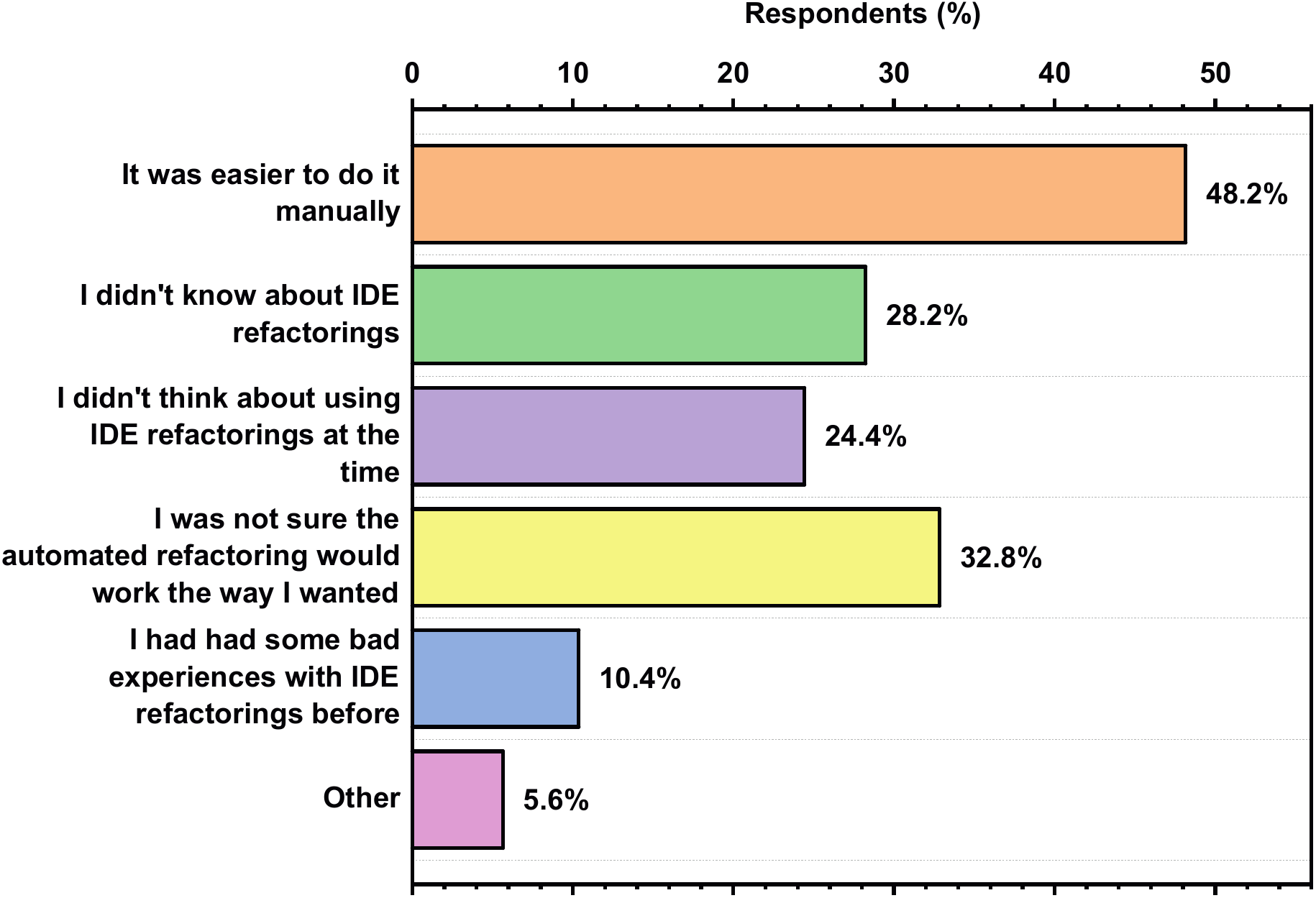}
  \caption{The answers to \textbf{Question 10}: \textit{In these scenarios (Inlining, Changing the Signature, Pulling/Pushing), what were your main reasons for not using the IDE refactoring feature?} (Out of 868 respondents)}
  \label{fig:question10}
\end{subfigure}
\vspace{-0.3cm}
\caption{Answers to \textbf{Questions 7--10} about different ways of using refactorings.}
\label{fig:questions7_10}
\vspace{-0.3cm}
\end{figure*}

The next section in the survey aimed to analyze how developers execute their intended refactoring, and the degree of their reliance on the IDE tools to do that. 
The results of survey questions corresponding to this investigation are presented in Figure~\ref{fig:questions7_10}.

The first question (\textbf{Question 7}) aimed to verify whether developers refactor code using built-in automated tools of IDEs or they perform it manually. First, we clustered refactorings into 3 main categories, namely the \textit{Rename}, the \textit{Extract}, and the \textit{Move} categories. Each category can target various code entities (\textit{e.g.}, rename class, method, attribute, etc.). Then, we designed our question to let respondents choose for each refactoring category whether they refactor their code using the appropriate IDE feature, or they rely on intuitive programming practices, such as \textit{Copy and Paste} or \textit{Find and Replace}. Since developers may select more than one choice, we report the percentage of developers selecting each option for each refactoring category. All percentages are showcased in Figure~\ref{fig:question7}.

In general, it can be seen that a significant number of respondents use IDE features to perform refactorings. \textit{Rename} is in the big lead, with as much as 85.8\% of participants saying that they used the IDE feature for it. This is a strong indicator that \textit{Rename} refactorings are implemented well as an IDE feature, their usage is intuitive, and they produce stable, reliable output. However, since renaming is a relatively simple change (at least compared to other refactorings), almost half the respondents also used the \textit{Find and Replace} feature to locate all instances of a given code element and rename it. Also, \textit{Rename} can be seen as the most universal of all the studied refactorings, because virtually no one said that they did not encounter a scenario for its use.

As for the \textit{Extract} refactoring, significantly fewer people used the IDE feature, about 54.7\%, especially when compared to the \textsc{Rename} feature. However, the \textsc{Extract} feature is still solicited by more than half of the respondents. 
Also, 10.7\% of the respondents indicated that they did not have a scenario where they had to extract something. These findings are in line with the recent study of Alcocer et al.~\cite{alcocer2020improving} where the authors reported the existence of certain usability issues of the \textit{Extract Method} refactoring when using IntelliJ IDEA.

Finally, \textit{Moving} code appears to be the first refactoring where the use of the IDE feature is not the most popular answer. 38.6\% of developers answered that they use the IDE feature and more than half the respondents (57.5\%) answered that they simply \textit{Copy and Paste}. Occasionally, moving elements can be basic, and in this case, simply copying and pasting suffices. However, when the move is more difficult, when it involves dependencies and complex relationships between objects, then the IDE feature might be simpler to use. Still, in general, \textit{Move} refactorings are almost as popular as \textit{Renames}, with only 4.4\% of participants saying that they did not encounter this scenario. 

It is also alarming to notice that for all three refactorings, about one-third of respondents performed the refactoring by manual editing (typing).

In the next \textbf{Question 8}, we selected all respondents who answered \textit{Used Find and Replace}, \textit{Used Copy and Paste}, or \textit{Edited manually} for at least one refactoring (\textbf{1,014 respondents}), and asked them about the reasons for not using the IDE refactoring features. The results are presented in Figure~\ref{fig:question8}.

Two answers are the most popular. 48.4\% of respondents said that they were not sure that the automated refactoring would work the way they wanted. This falls in line with the popular notion that developers often do not trust automated refactoring tools~\citep{brant2015refactoring,bogart2020increasing,silva2016we,murphy2011we,vakilian2012use}. Also, 45.6\% of respondents said that certain refactorings were easier to conduct manually. This puts the results of the previous question into perspective: a lot of developers use ``manual'' ways of conducting refactorings, because sometimes it is just more straightforward to do so. Therefore, one of the main takeaways for our IDE development teams is the importance of making the refactoring tools simpler and more intuitive. Since the \textsc{Rename} feature is the most successful one in terms of adoption, analyzing how developers activate it would help in understanding its success, and potentially replicating it to other types of refactorings. 

Fewer participants selected other options: 24.7\% said that they did not think about using IDE refactoring tools at the time, 21.7\% said that they did know about the IDE tools, and 15.9\% said that they had certain negative experiences with refactoring features before. When given a prompt to answer freely, some respondents mentioned several issues with refactoring features: performance issues, namespace confusion, difficulties with refactorings that involve significant changes to the logic of the program. This feedback is critical to our IDE development teams, and therefore, there will be a follow up with these respondents to seek more technical details with regard to their issues.

After this, we repeated the same line of questioning in regards to three other refactorings, namely, \textit{Inlining} a variable or a method, \textit{Changing the signature} of a function, and \textit{Moving a method up and down the class hierarchy} (known also as \textit{Pull up} and \textit{Push down}). The answers to \textbf{Question 9} are demonstrated in Figure~\ref{fig:question9}.

\begin{figure*}
  \centering
  \vspace{0.7cm}
  \includegraphics[width=\textwidth]{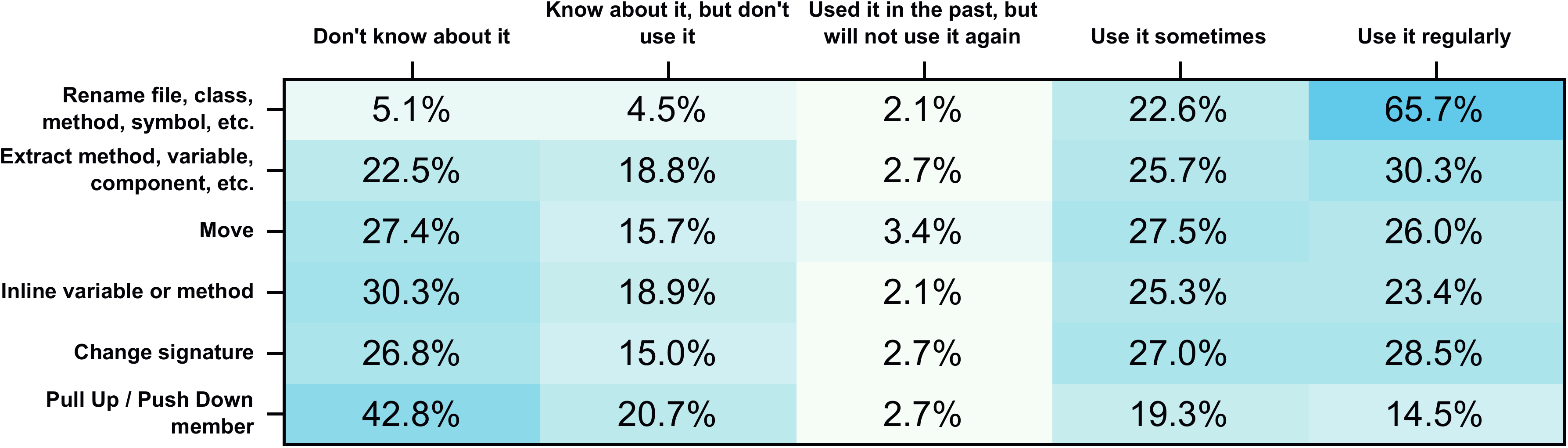}
  \vspace{-0.2cm}
  \caption{The answers to \textbf{Question 11}: \textit{How familiar are you with the following IDE refactoring features?} (Out of 1,167 respondents)}
  \label{fig:question11}
  \vspace{-0.3cm}
\end{figure*}

It can be seen that \textit{Changing the signature} is approximately as popular as \textit{Extracting} in \textbf{Question 7}, with 10.6\% of respondents saying that they did not recently have this scenario in their work and 53.6\% saying that they used automated IDE tools. For this refactoring, editing manually was also a very popular answer.
As for the other two refactorings, they seem to be less popular than the previous ones. For \textit{Inlining}, a quarter of the respondents said that they did not recently encounter this scenario in their work. Still, half the respondents said that they use IDE features for \textit{Inlining}.
\textit{Pulling up / Pushing down} appears to be the least popular of the chosen refactorings. 30\% of respondents did not recently encounter this refactoring in their work, and only 27\% used IDE features for these refactorings.

In \textbf{Question 10}, we once again selected participants who answered \textit{Used Find and Replace}, \textit{Used Copy and Paste}, or \textit{Edited manually} for at least one refactoring from \textbf{Question 9} (\textbf{868 respondents}) and asked them about their reasons for not using the IDE tools. The results are presented in Figure~\ref{fig:question10}. The general distribution of answers is similar to Figure~\ref{fig:question8}, but there are some differences between them. There are more answers \textit{It was easier to do it manually} and \textit{I did not know about IDE refactorings}. It is possible that these refactorings are less universally known. Another takeaway for our IDE development teams is to attract developers' attention to the possibility of using these refactorings, perhaps by implementing special tooltips or IDE notifications.

\vspace{0.1cm}

\observation{Renames are the most universal refactoring, with virtually everyone using them. 85.8\% of participants renamed elements using IDE tools. On the other side, 30\% of respondents stated that they did not recently perform \textit{Pull Up} and \textit{Push Down} refactorings. Despite the existence of IDE refactoring features, developers still manually refactor their code, including \textit{Find and Replace} for Rename, \textit{Copy and Paste} for Move, \textit{Editing manually} for Changing signature. Respondents justified their reluctance to use IDE refactoring features with not knowing the outcomes, along with manual refactoring being more intuitive.}

\subsection{IDE Refactoring Features}
\label{sec:IDEref}

In this section, we focused our questions specifically on IDE refactoring features. \textbf{Question 11} asked the developers about their familiarity with IDE refactoring features for the same six refactorings studied in Section~\ref{sec:approaches}. The heatmap with all the results is demonstrated in Figure~\ref{fig:question11}.

Our first observation from the figure is the middle column. For all the refactorings, there were equally very few respondents who said that they refactored in the past, but will not do it again. This is a very positive insight meaning that even with all the concerns raised about automatic refactoring features, developers generally do not give up on them.

Coming to the other answers, they correlate fairly well with the previous questions. Once again, we can see the same two main outliers. \textsc{Rename} is the most popular refactoring feature, with 65.7\% of respondents saying that they use it regularly and 22.6\% more saying that they use it sometimes. On the other hand, only 5.1\% of participants said they are unaware of the \textsc{Rename} IDE feature.
Meanwhile, \textsc{Pull up / Push down} remains the least popular: up to 42.8\% of developers did not know about the \textsc{Pull up / push down} IDE feature, only 19.3\% of respondents used it sometimes, and only 14.5\% used it regularly. 
All the other automated refactoring features (\textsc{Extract}, \textsc{Move}, \textsc{Inline}, and \textsc{Change signature}) are distributed more similarly between \textsc{Rename} and \textsc{Pull up / Push down}. 

\textbf{1,078} respondents (92.4\%) selected \textit{Use it sometimes} or \textit{Use it regularly} for at least one refactoring feature. For these participants, the next block of questions was unlocked.
In \textbf{Question 12}, we asked the participants about their overall experience when using the last several IDE refactoring features. The results are presented in Table~\ref{table:question12}.

\begin{table}[h]
\centering
\caption{The answers to \textbf{Question 12}: \textit{Please think about the last several times you used IDE refactorings. How happy were you with the overall experience?} (Out of \textbf{1,078} respondents)}
\label{table:question12}
\begin{tabular}{cc}
\toprule
\textbf{Answer} & \textbf{\% of respondents} \\
\midrule
Not at all happy & 0.3\% \\
Not happy & 1.6\% \\
Neither happy nor unhappy & 12.6\% \\
Happy & 56.9\% \\
Very happy & 28.6\% \\
\bottomrule
\end{tabular}
\end{table}

It can be seen that the overall experience is overwhelmingly positive, with 85.5\% of developers giving positive responses, and as much as 28.6\% saying that they were \textit{Very happy} with the last several uses of IDE refactoring features.

Still, even though only 1.9\% of developers answered either \textit{Not happy} or \textit{Not at all happy}, there were 12.6\% respondents who answered \textit{Neither happy nor unhappy}. It is very important to focus on developers who do not give positive answers to get a deeper understanding of underlying shortcomings in IDE refactoring features. To do this, in \textbf{Question 13}, we asked all the respondents who did not give a positive answer to share their experience in an open form. Several issues were brought up by several participants:

\begin{enumerate}
    
    \item 22 developers mentioned that their main problem was a certain negative experience when a refactoring was performed inaccurately. 6 developers specifically mentioned that a refactoring broke the code or introduced new errors. 

    \item 7 developers mentioned that using an IDE refactoring tool might be slower than doing the refactoring manually. On the other hand, for very large projects, the problem of correctly taking care of all namespaces for various variables and functions becomes more difficult. 
    
    \item 3 developers mentioned that IDE refactoring tools have a steep learning curve for them. As seen in previous questions, almost all developers understand what \textsc{Rename} is, but may never use some more specific features such as \textsc{Pull up / Push down} simply because they have no environment to understand what they do and what they are for. 5 more developers mentioned that the existing features that they know are confusing for them.
    
    \item Related to the last point, 5 developers mentioned that they would want a more direct and more visible representation of IDE refactoring tools: special tips, or maybe a notification when a refactoring is possible.
    
    \item 3 developers mentioned a general lack of trust in all automated tools.
    
    \item 2 developers brought up another interesting issue: applying an automated refactoring can violate a specific coding style guideline or formatting. Therefore, even if the refactoring is done correctly, it might still need some editing afterward.
\end{enumerate}

This list is not comprehensive, but it can serve as a blueprint to get an idea of the current challenges that developers are facing with regard to the usage of automated IDE refactoring tools.

Similar to other IDE features, refactorings can be reverted in case their outcome seems unexpected to the developer. 
To target this event, we asked \textbf{Question 14} about how often developers had to undo a refactoring performed by the IDE. In order for the answers not to be vague, we specified the approximate percentage of time for each answer. The results are presented in Table~\ref{table:question14}. 

\begin{table}[h]
\centering
\caption{The answers to \textbf{Question 14}: \textit{How often do you undo or revert an IDE refactoring action because you’re unhappy with the result?} (Out of \textbf{1,091} respondents)}
\label{table:question14}
\begin{tabular}{cc}
\toprule
\textbf{Answer} & \textbf{\% of respondents} \\
\midrule
Often ($>$75\% of the time) & 3.0\% \\
Every other time ($\approx$50\% of the time) & 6.2\% \\
Sometimes ($\approx$25\% of the time) & 32.3\% \\
Rarely ($\approx$5\% of the time) & 50.8\% \\
Never & 7.7\% \\
\bottomrule
\end{tabular}
\end{table}

Naturally, it can be seen that in this question, extreme answers were rare. Indeed, only 7.7\% of respondents said that they \textit{Never} revert, but at the same time, only 6.2\% said that they do it \textit{Every other time ($\approx$50\% of the time)} and only 3\% said that they do it \textit{Often ($>$75\% of the time)}. The vast majority of our respondents lied in between these poles, with 32.3\% saying they undo refactoring actions \textit{Sometimes ($\approx$25\% of the time)} and half of all participants (50.8\%) saying they do it \textit{Rarely ($\approx$5\% of the time)}. 

\begin{figure}[t]
  \centering
  \vspace{-0.1cm}
  \includegraphics[width=\columnwidth]{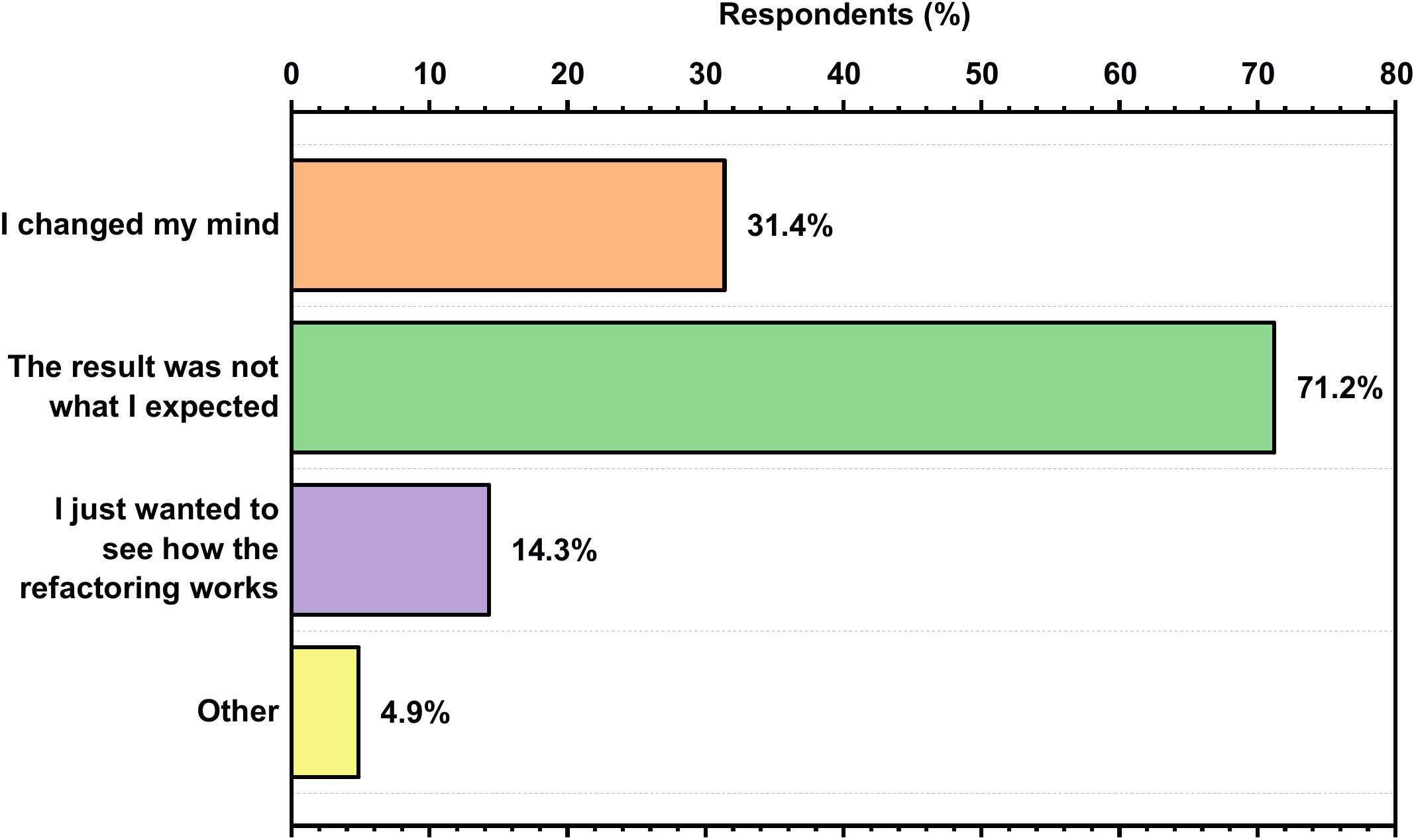}
  \caption{The answers to \textbf{Question 15}: \textit{The last few times you undid or reverted an IDE refactoring feature, what were the reasons?} (Out of \textbf{1,007} respondents)}
  \label{fig:question15}
  \vspace{-0.2cm}
\end{figure}

In general, \textsc{Undo} is one of the most fundamental actions in programming, it happens all the time for various reasons. In a way, it is an inherent part of the creative process. Nevertheless, it is captivating for us to discover the reasons behind developers undoing IDE refactoring actions. We asked this in \textbf{Question 15} of all the respondents who did not choose \textit{Never} in \textbf{Question 14} (\textbf{1007} respondents). The answers are presented in Figure~\ref{fig:question15}.

The most popular answer by far is that the produced result of the IDE feature was not what the developer expected, with 71.2\% of participants selecting this option. 31.4\% of respondents said that they changed their mind, and also 14.3\% said that they just wanted to see how the refactoring would work. Among other reasons that the developers shared were their own mistakes, mishaps, and wrong configuration of refactoring (like namespaces of formats). 

\vspace{0.1cm}
\observation{85.5\% of the respondents said that their experience with the IDE refactoring features is positive. A deeper inquiry about the existing challenges resulted in a list of possible issues to consider, including the difficulty of refactoring large projects, and the learning curve of some refactoring types. The vast majority of developers occasionally undo or revert refactorings, with the main reason being that the refactoring produced the result that was unexpected.}

\subsection{Previewing Refactorings}

Previewing the outcome of refactoring before its execution is one of the main features of IntelliJ IDEs.
Previously, in \textbf{Question 13}, some developers mentioned that their reliance on the refactoring feature highly depends on the preview, and that for some of them, the preview makes the biggest difference. We decided to investigate this further, so our next block of questions is aimed specifically at the process of previewing the refactoring.

\begin{figure}[h]
  \centering
  \includegraphics[width=2.65in]{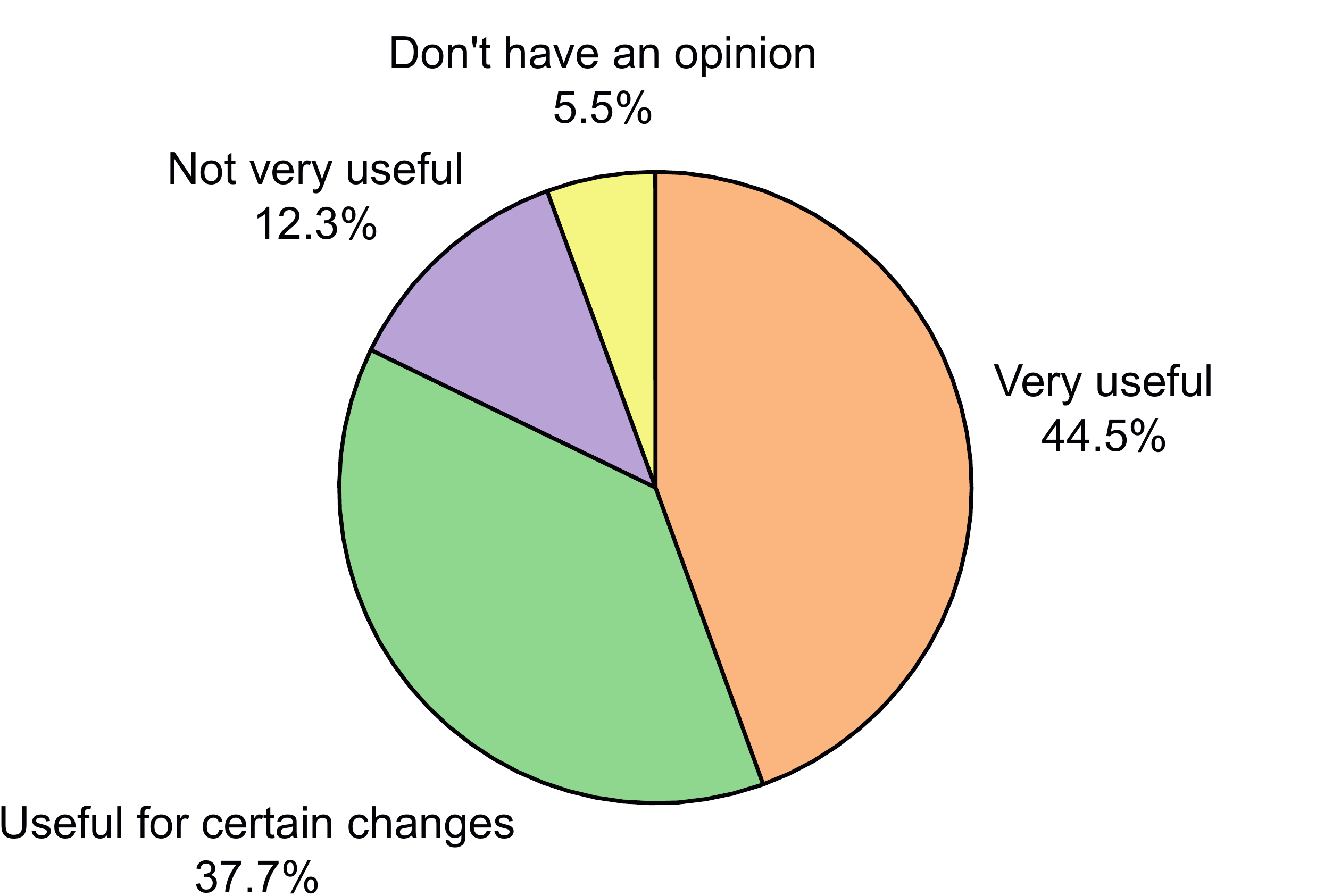}
  \caption{The answers to \textbf{Question 16}: \textit{When refactoring the code, do you find it useful to preview all of your changes before applying them?} (Out of \textbf{1,167} respondents)}
  \label{fig:question16}
  \vspace{-0.2cm}
\end{figure}

In \textbf{Question 16}, we asked all the respondents whether they find the previewing useful. The distribution of the results is presented in Figure~\ref{fig:question16}.
The opinion about previewing refactorings is also largely positive. 44.5\% of the respondents said that the previews are \textit{Very useful}, and 37.7\% said that they are \textit{Useful for certain changes}. Overall, this constitutes more than 80\% of participants. On the other hand, 12.3\% said that the previews are \textit{Not very useful} and 5.5\% more did not have an opinion on the subject.

While this in itself proves the importance of the previewing function, it is of interest to dive deeper into both sides of this question. For this reason, we divided the respondents into those who gave positive answers and those who did not, and asked them separate questions.

For participants who answered positively, \textbf{Question 17} asked for specific types of refactorings, for which they find the preview to be especially useful.
The distribution of answers correlated well with the overall distribution of the popularity of the refactoring feature usage (as indicated by the last two columns in Figure~\ref{fig:question11}), so it might be the case that the preview is used more or less similarly for different refactorings, at least no obvious anomalies were present.

Then, in \textbf{Question 18}, we asked the respondents about their reasons for wanting to see the preview. The most popular answer, indicated by as many as 227 developers, is making sure that everything is done as they want it to be, with 22 developers specifically saying they are making sure that the code does not break, and 13 developers mentioning using this to combat a distrust in automatic features. However, other specific aspects were brought up.

\begin{enumerate}
    \item 78 developers specifically mentioned making sure that the IDE will not do anything extra and checking the impact of the changes on the whole codebase. Often, this has to do with renaming, and developers check namespaces, imports, and occurrences in comments.
    \item 8 developers check the preview carefully because they say it prevents them from wasting time later to check the Version Control System (VCS) diff or undo the unwanted changes.
    \item 5 developers said that they check the readability of the code during the change, making sure that applying refactoring will not worsen it, as well as check code style. They mentioned that in this regard, the visual aspect of the preview is very important.
    \item One developer mentioned that they are interested in the preview as a means to comprehend their code. Basically, they look at the preview not to check if the IDE missed something, but rather to check whether \textit{they} forgot something. Several respondents mentioned that the preview allows them to think about the necessity of the change one more time.
\end{enumerate}

It can be seen that apart from its main purpose, the preview sometimes serves different other goals. We also observed that the reasons for seeing the preview correlate well with the general shortcomings of refactoring features that developers mentioned in Section~\ref{sec:IDEref}. The preview can elevate many of the concerns that have to do with refactorings, even such specific ones as, for example, the uniformity of code style.

\begin{figure}[h]
  \centering
  \vspace{-0.2cm}
  \includegraphics[width=\columnwidth]{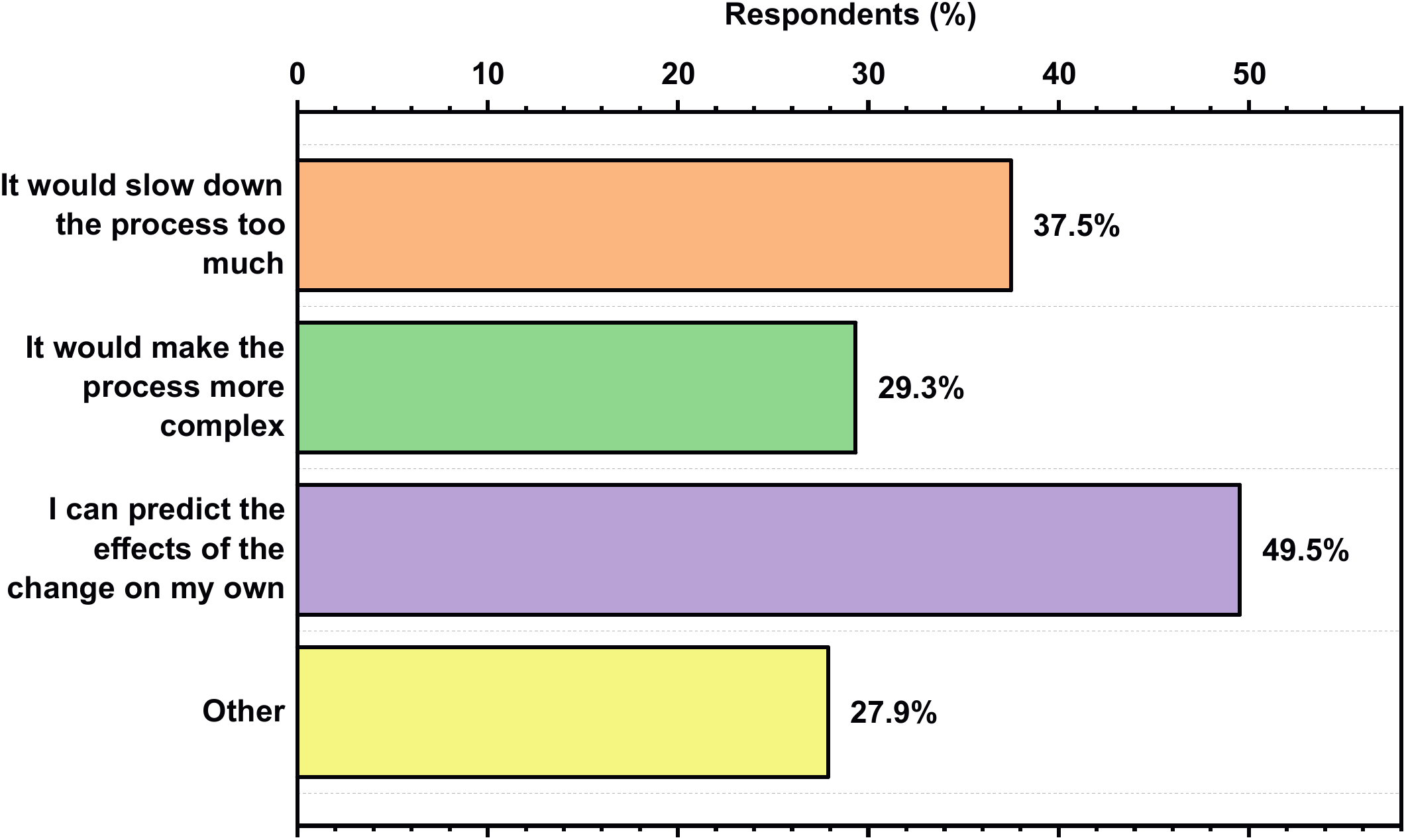}
  \caption{The answers to \textbf{Question 19}: \textit{What are your main reasons for not finding a preview useful?} (Out of \textbf{208} respondents)}
  \label{fig:question19}
  \vspace{-0.2cm}
\end{figure}

It is also important to understand the reasoning of participants who did not answer positively to the question about the usefulness of the preview. To such respondents, we showed \textbf{Question 19} asking them for their reasons. The distribution of the results is shown in Figure~\ref{fig:question19}.

It can be seen that almost half of the respondents who were asked this question selected the answer \textit{I can predict the effects of the change on my own}. The other responses are a little less popular: the previewing slowing down the IDE too much or adding too much to the complexity of the process.

When given a prompt to write the \textit{Other} answer, respondents mentioned several reasons.

\begin{enumerate}
    \item 13 developers said that they prefer to use the VCS diff post-factum, motivating it with the fact that it produces the same functionality but also merges it with the possibility to verify a wider set of changes.
    \item 10 developers were simply very different from the ones in the previous questions, saying that they would rather simply run the code, and if it fails, they can always undo the change. 
    \item 6 developers mentioned that the preview function is too complex for them or confuses them.
    \item 5 developers did not know about or did not encounter the preview feature.
\end{enumerate}

In general, it seems that the opinion strongly depends on the developers, and in our survey, more developers demonstrated a cautious approach.

\vspace{0.1cm}

\observation{Previewing is a critical feature for the IDE refactoring process. 82.2\% of our respondents found it useful at least for specific changes. Apart from simply making sure that the refactoring is what is intended, it is useful in other situations, such as a fail-safe against a fast decision, a reason to review the change, as well as a verification for code readability and compliance with the code style of the project. A minority of our respondents were not positive about the preview, with half of them justifying this with being able to predict the refactoring outcome without the preview.}

\subsection{Final Thoughts}

The last question in our survey, \textbf{Question 20}, was an open-ended question for describing general thoughts about refactorings and IDE refactoring features. In general, the respondents' messages were positive. 177 developers expressed their compliments, respondents very often mentioned that refactorings are a cornerstone of the development process, and for some of them --- the main reason to use an IDE in the first place. Let us enumerate the key findings in their feedback.

\textbf{Refactoring knowledge.} 52 developers reiterated a previous point about refactoring tools having a steep learning curve. They suggested that refactorings should be directly enabled in IDEs, through tips and popups. Some developers even suggested having interactive tutorials within the IDE that show up for first-time usage. Interestingly, several respondents mentioned that they are quite familiar with some refactoring types and hardly know anything about others.

\textbf{Language difference.} 9 developers commented on how refactorings behave differently in different languages. The biggest difference is noted between statically typed languages like Java and dynamically typed languages like Python, with refactorings in the latter ones being much less predictable. This is understandable from the nature of the language itself, however, tool developers should pay closer attention to refactorings in dynamically typed languages. This supports the findings in other works~\cite{pinto2013programmers,Newman:2018:SDP:3242163.3242170} observing that developers are interested in refactorings for more languages, including the dynamically typed ones.

\textbf{Scope.} As mentioned before, one common problem the respondents have with refactorings is how they treat complex namespaces. Developers also mentioned the ability for any refactoring to be easily undone as a useful feature. Furthermore, developers suggested a more explicit and clear way to mark certain source files or code elements as a no-go zone for refactorings.

\textbf{Complex features.} Finally, several developers mentioned their thoughts on the complexity of tasks that refactorings solve. Some developers expressed the idea that most of the automatic refactorings, like the ones we study in this survey, are useful and time saviors, but they represent basic atomic actions. They suggested supporting more complex features like splitting a large complex method or class into several smaller ones.
On the other hand, several developers mentioned that for simpler changes, it is often much faster to apply the refactorings manually, and that automated refactorings do not need to strive to replace the simpler changes completely.

\vspace{-0.1cm}
\section{Threats to Validity}

A large percentage of our participants (48.1\%) have more than 10 years of professional experience.
Moreover, the respondents might be more experienced with refactorings because IntelliJ-based IDEs are known for their rich support of refactorings. In our questionnaire, we focused only on a part of refactorings the IntelliJ-based IDEs support, which are the most well-known and studied~\cite{murphy2011we}.
However, in the \textit{Final Thoughts} section of the questionnaire, the respondents were free to provide feedback about their experience with any refactoring. 

Our survey involved participants who use a large variety of languages and IDEs, which could influence the way they plan their refactorings.
We did not classify our findings by a specific language or an IDE.
Further research aimed at investigating the possible differences is necessary. 

Since all participants in our survey are paid subscribers of JetBrains IDEs, 
our results may not generalize to other contexts and other companies. Also, concerning the correctness of our interpretation of open responses, we did not discuss all responses because some of them are open to various interpretations, and follow-up surveys or interviews are needed to clarify them.

\vspace{-0.2cm}
\section{Implications}

\textbf{Educating about refactorings.}
The results show that developers are not familiar with some of the refactoring types that IDEs support (see Figure~\ref{fig:question11}).
Also, some developers might be cautious about the side-affects of refactoring, so they are not likely ready to perform any refactoring unless they get to know how it works (see Figure~\ref{fig:question8}).
To encourage users to use automatic refactorings tools in IDEs, it might be useful to start with educating users about them.
For example, along with showing a possibility to perform a refactoring, IDEs could provide some information about its purpose as well.
The knowledge about how refactoring works will help IDEs to respond to user expectations (see Table~\ref{table:question14} and Figure~\ref{fig:question15}).

\textbf{Increasing awareness about refactoring possibilities.}
One of the reasons the developers do refactorings manually is that they do not know about the possibility to perform them automatically~\cite{silva2016we}.
It could be helpful if IDEs suggested possible refactoring opportunities for the user. 
The results show that users often refactor their code, sometimes for more than an hour, so automatic refactoring recommendations could save them time.

\textbf{Supporting complex refactorings.}
Since some developers expressed their need in support of complex refactorings in IDEs, it would be exciting to support batch refactorings \cite{sousa2020characterizing,bibiano2019quantitative}, even if they are less intuitive. Such series of transformations would result in more atomic methods that better optimize structural metrics, such as complexity and lines of code. A lot of research efforts have focused on exploring the impact of refactoring on software quality using metrics~\cite{mkaouer2015many,alomar2019impact,cedrim2017understanding,pantiuchina2018improving}, and it would be interesting to turn this into actual IDE features.

\vspace{-0.2cm}

\section{Conclusion}

In this paper, we presented the results of a large-scale survey about refactoring usage, conducted by JetBrains Research among the users of IntelliJ-based IDEs. The survey consists of several blocks of questions: familiarity with refactorings, using different approaches to conduct refactorings, making use of specific IDE refactoring features, and using previews when employing refactorings. Together with the background information at the start and some final thoughts at the end, the results of this survey represent a unique perspective of 1,183 experienced developers.

We hope that our results can be useful for both researchers and practitioners, and the presented opinions of more than a thousand developers can be used to perfect our methods and our tools.

\section*{Acknowledgements}

We would like to thank the Product Management team and the Market Research and Analytics team of JetBrains for their advice and guidance during the designing and conducting of the survey.

\bibliographystyle{ACM-Reference-Format}
\bibliography{cite}

\end{document}